\pdfoutput=1
\RequirePackage{ifpdf}
\ifpdf 
\documentclass[pdftex]{sigma}
\else
\documentclass{sigma}
\fi

\numberwithin{equation}{section}

\def\C{{\mathbb C}}
\def\Z{{\mathbb Z}}

\def\F{{\mathbb F}}
\def\R{{\mathbb R}}
\def\P{{\mathbb P}}
\def\A{{\mathbb A}}

\begin{document}

\renewcommand{\PaperNumber}{054}

\FirstPageHeading

\ShortArticleName{Discrete Integrable Equations over Finite Fields}

\ArticleName{Discrete Integrable Equations over Finite Fields}

\Author{Masataka KANKI~$^\dag$, Jun MADA~$^\ddag$ and Tetsuji TOKIHIRO~$^\dag$}
\AuthorNameForHeading{M.~Kanki, J.~Mada and T.~Tokihiro}

\Address{$^\dag$~Graduate School of Mathematical Sciences, University of Tokyo,\\
\hphantom{$^\dag$}~3-8-1 Komaba, Tokyo 153-8914, Japan}
\EmailD{\href{mailto:kanki@ms.u-tokyo.ac.jp}{kanki@ms.u-tokyo.ac.jp}, \href{mailto:toki@ms.u-tokyo.ac.jp}{toki@ms.u-tokyo.ac.jp}}

\Address{$^\ddag$~College of Industrial Technology, Nihon University,\\
\hphantom{$^\ddag$}~2-11-1 Shin-ei, Narashino, Chiba 275-8576, Japan}
\EmailD{\href{mailto:mada.jun@nihon-u.ac.jp}{mada.jun@nihon-u.ac.jp}}

\ArticleDates{Received May 18, 2012, in f\/inal form August 15, 2012; Published online August 18, 2012}

\Abstract{Discrete integrable equations over f\/inite f\/ields are investigated.
The indeterminacy of the equation is resolved by treating it over a f\/ield of rational functions instead of the f\/inite f\/ield itself.
The main discussion concerns a generalized discrete KdV equation related to a Yang--Baxter map.
Explicit forms of soliton solutions and their periods over f\/inite f\/ields are obtained.
Relation to the singularity conf\/inement method is also discussed.}

\Keywords{integrable system; discrete KdV equation; f\/inite f\/ield; cellular automaton}

\Classification{35Q53; 37K40; 37P25}

\section{Introduction}
\label{sec1}

Cellular automata are discrete dynamical systems which provide simple and ef\/f\/icient tools for modeling complex phenomena \cite{Wolfram}.
Since each cell of a cellular automaton takes only a f\/inite number of states, it seems natural to
describe its time evolution by utilising a f\/inite f\/ield.
In particular, if we can construct f\/inite f\/ield analogues of dynamical equations whose mathema\-ti\-cal structures are well studied,
such as integrable systems, this construction may give some fundamental methods for analysing models of cellular automata.
Discrete analogues of integrable equations have been widely investigated, however,
their extension over a f\/inite f\/ield  has less been examined.
One of the reasons for this may be that the time evolution of a nonlinear system is not always well def\/ined over a f\/inite f\/ield.
For example, let us consider the discrete KdV equation
\begin{gather}
\frac{1}{x_{n+1}^{t+1}}-\frac{1}{x_n^t}+\frac{\delta}{1+\delta}\left(x_n^{t+1}-x_{n+1}^t \right)=0,
\label{dKdV1}
\end{gather}
over a f\/inite f\/ield $\F_q$ where $q=p^m$, $p$ is a prime number and $m\in\Z_{+}$. Here $n,t \in \Z$ and $\delta$ is a~parameter.
If we put
\[
\frac{1}{y_n^t}:=(1+\delta)\frac{1}{x_n^{t+1}}-\delta x_n^t
\]
we obtain equivalent coupled equations
\begin{gather}
x_n^{t+1} =\dfrac{(1+\delta)y_n^t}{1+\delta x_n^ty_n^t},\qquad
y_{n+1}^{t} =\dfrac{(1+\delta x_n^ty_n^t)x_n^t}{1+\delta}.
\label{dKdV2}
\end{gather}
Clearly \eqref{dKdV2} does not determine the time evolution when $1+\delta x_n^t y_n^t\equiv 0$.
Over a f\/ield of characteristic~0 such as~$\C$, the time evolution of $(x_n^t,y_n^t)$ will not hit this exceptional line
for generic initial conditions, but on the contrary, the evolution comes to this exceptional line in many cases over a f\/inite f\/ield as a division by $0$ appears.

A pioneering work on integrable equations over f\/inite f\/ields is that by Doliwa, Bia{\l}ecki and Klimczewski \cite{BD, DBK}.
They used an algebro-geometric approach to construct soliton solutions to discrete integrable equations over f\/inite f\/ields in Hirota's bilinear
form.
For the discrete KdV equation, the bilinear form is written as
\begin{gather}
(1+\delta)\sigma_{n+1}^{t+1}\sigma_n^{t-1}=\delta \sigma_{n+1}^{t-1}\sigma_n^{t-1}+\sigma_n^t\sigma_{n+1}^t.
\label{bilineardKdV}
\end{gather}
The $N$-soliton solution to the equation \eqref{bilineardKdV} is given as
\begin{gather}
\sigma_n^t=\det_{1\le i,j\le N}\left( \delta_{ij}+\frac{\gamma_i}{l_i+l_j-1}\left(\frac{1-l_i}{l_i}\right)^t
\left(\frac{l_i+\delta}{1+\delta-l_i}\right)^n\right)
\label{Nsoliton}
\end{gather}
where $\gamma_i,\ l_i$ $(i=1,2,\dots ,N)$ are arbitrary parameters satisfying $l_i \ne l_j$ for $i \ne j$.
Hence if we choose $l_i \not\equiv 0, 1+\delta$ $(i=1,2,\dots ,N)$ and $l_i+l_j \not\equiv 1$ $(1\le i, j \le N)$, the $N$ soliton
solution~\eqref{Nsoliton} is well def\/ined for all $(n, t) \in \Z^2$ and gives a time evolution pattern over a f\/inite f\/ield.
A~similar approach was also used for discrete KP equation in the bilinear form~\cite{BN}.
However, since the nonlinear form of the discrete KdV equation~\eqref{dKdV1} is obtained from \eqref{bilineardKdV} by
putting $x_n^t:=\frac{\sigma_n^t\sigma_{n+1}^{t-1}}{\sigma_{n+1}^t\sigma_n^{t-1}}$, well def\/ined $N$-soliton solutions $x_n^t$ of~\eqref{dKdV1} or~\eqref{dKdV2} cannot be obtained from~\eqref{Nsoliton} because $x_n^t$ is not def\/ined if $\sigma_{n+1}^t\equiv 0$ or $\sigma_n^{t-1} \equiv 0$.
Indeterminacy of the time evolution for a~generic initial state cannot be avoided either when we use Hirota's bilinear form.
Note that if we consider the equation over $\P\F_q:=\F_q\cup \{\infty\}$ instead of $\F_q$ by adding a value $\infty$, we frequently hit the indeterminate values $\frac{0}{0}$, $\infty + 0$, $0 \cdot \infty $ and so on, which causes further problems.

In this article, we propose a prescription to determine the time evolution of a nonlinear system over f\/inite f\/ields by taking as examples the discrete KdV equation \eqref{dKdV1} and its generalization. We show in Section~\ref{section2} that the initial value problem is well def\/ined and we investigate $N$-soliton solutions in Section~\ref{section3}. The last section is devoted to discussing the relation of our method to the singularity conf\/inement method~\cite{Grammaticosetal}.

\section{A generalized discrete KdV equation over a function f\/ield}\label{section2}

\subsection{Discrete KdV equation}

First we explain how the indeterminate values appear through the time evolution by examining the discrete KdV equation \eqref{dKdV2} over $\F_7:=\{0,1,2,3,4,5,6\}$.
If we take $\delta=1$, \eqref{dKdV2} turns into
\[
x_n^{t+1}=\dfrac{2y_n^t}{1+x_n^ty_n^t},\qquad
y_{n+1}^t=\dfrac{(1+x_n^ty_n^t)x_n^t}{2}.
\]
Suppose that $x_1^0=6$, $x_2^0=5$, $y_1^0=2$, $y_1^1=2$, then
we have
\[
x_1^{1}=\frac{4}{13} \equiv 3,\qquad y_2^{0}=\frac{78}{2} \equiv 4 \ \mod 7.
\]
With further calculation we have
\[
x_1^2=\frac{4}{7} \equiv \frac{4}{0},\qquad y_2^1=\frac{21}{2} \equiv 0,\qquad x_2^1=\frac{8}{21} \equiv \frac{1}{0}.
\]
Since $\frac{4}{0}$ and $\frac{1}{0}$ are not def\/ined over $\F_7$, we now extend $\F_7$ to $\P\F_7$ and
take $\frac{j}{0}\equiv \infty$ for $j\in\{1,2,3,4,5,6\}$.
However, at the next time step, we have{\samepage
\[
x_2^2=\frac{2 \cdot 0}{1+ \infty \cdot 0},\qquad y_3^1=\frac{(1+ \infty \cdot 0)\cdot \infty}{2}
\]
and reach a deadlock.}

\begin{figure}[t]
\centering
\includegraphics[width=9.5cm]{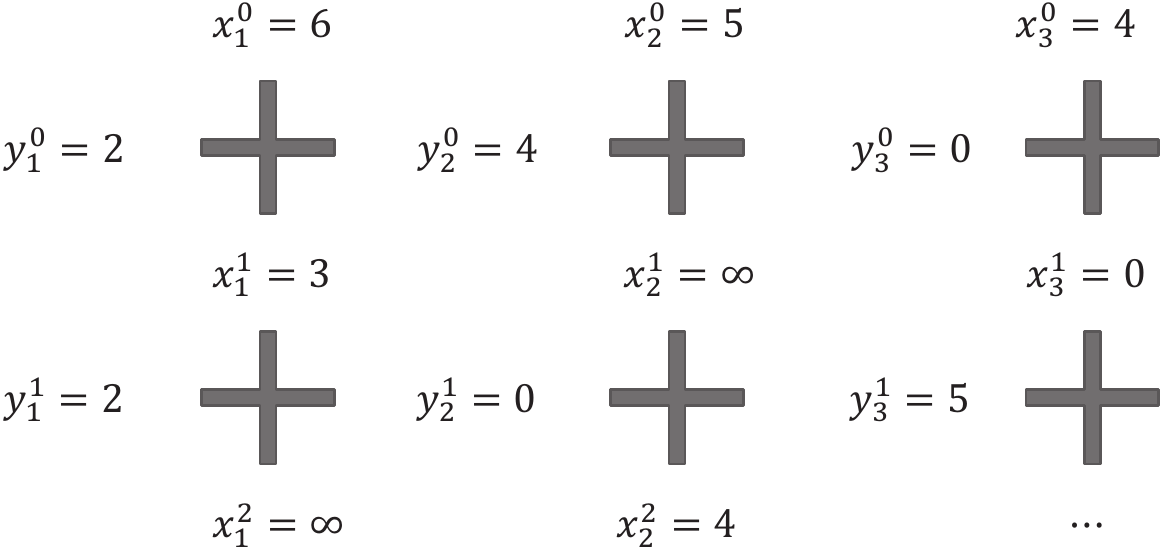}
\caption{An example of the time evolution of the coupled discrete KdV equation \eqref{dKdV2} over $\P\F_7$ where $\delta=1$.}
\label{figure1}
\end{figure}

\begin{figure}[t!]
\centering
\includegraphics[width=10cm]{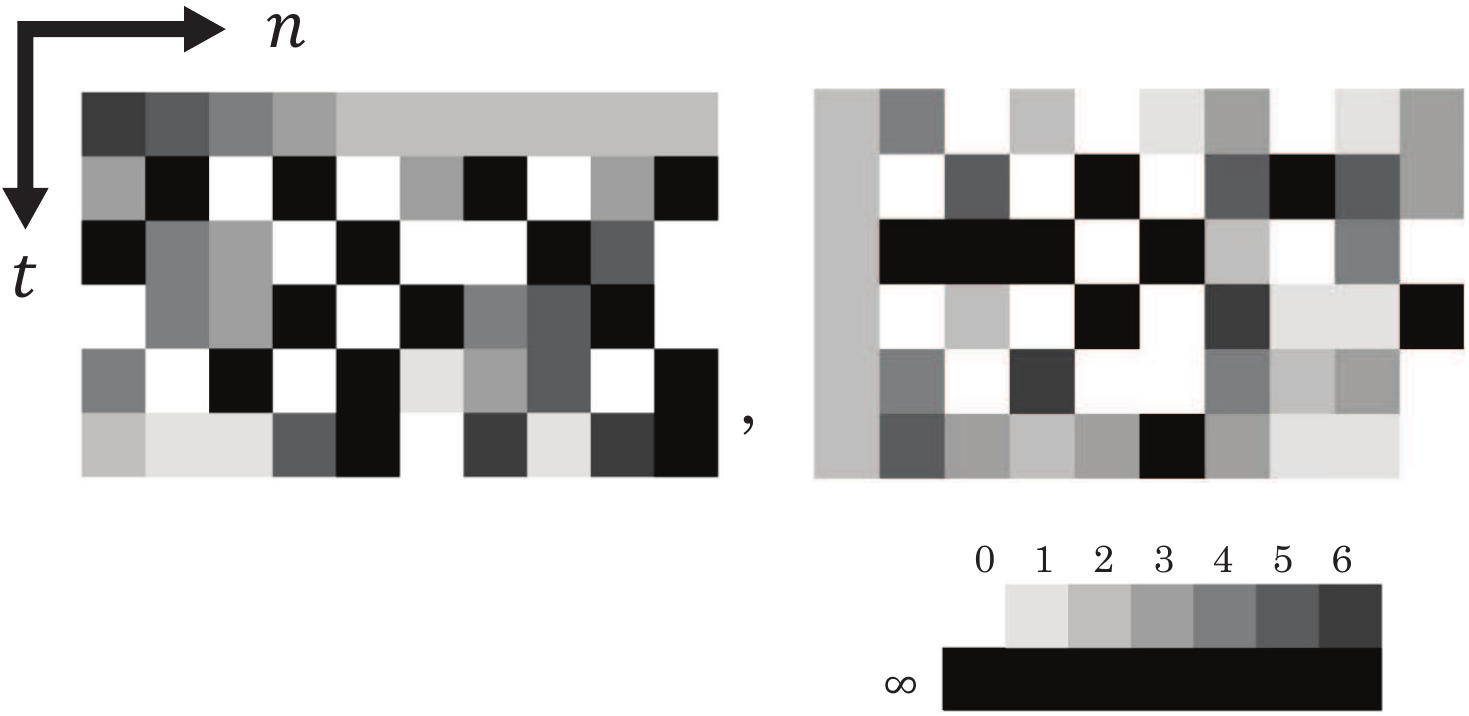}
\caption{The time evolution pattern of $x_n^t$ (left) and $y_n^t$ (right) of \eqref{dKdV2} over $\P\F_7$ where $\delta=1$. Elements of $\P\F_7$ are represented on the following grayscale: from $0$ (white) to $6$ (gray) and $\infty$ (black). See the scale bar in the f\/igure.}
\label{figure2}
\end{figure}

Therefore we try the following two procedures:
[I] we keep $\delta$ as a parameter for the same initial condition, and obtain as a system over $\F_7(\delta)$,
\begin{gather*}
x_1^{1} =\frac{2(1+\delta)}{1+5\delta},\quad y_2^{0}=\frac{6(1+5\delta)}{1+\delta},\qquad
x_2^1 =\frac{6(1+\delta)(1+5\delta)}{1+3\delta+3\delta^2},
\qquad y_2^1=\frac{2(1+2\delta+4\delta^2)}{(1+5\delta)^2},\\
x_1^2=\frac{2(1+\delta)(1+5\delta)}{1+2\delta+4\delta^2},\qquad
x_2^2 =\frac{4(1+\delta)(2+\delta)(3+2\delta)}{(1+5\delta)(5+5\delta+2\delta^2)},\qquad y_3^1=\frac{2(5+5\delta+2\delta^2)}{(2+\delta)^2}.
\end{gather*}
[II] Then we put $\delta=1$ to have a system over $\P\F_7$ as
\begin{gather*}
x_1^{1} =3,\qquad y_2^{0}=4,\qquad x_2^1=\frac{72}{7}\equiv \infty,
\quad y_2^1=\frac{14}{36}\equiv 0,\qquad x_1^2=\frac{24}{7} \equiv \infty,\\
x_2^2 =\frac{120}{72} \equiv 4,\qquad y_3^1=\frac{24}{9} \equiv 5.
\end{gather*}
Thus all the values are uniquely determined over~$\P\F_7$.
Figs.~\ref{figure1} and~\ref{figure2} show a time evolution pattern of the discrete KdV equation~\eqref{dKdV2} over~$\P\F_7$ for the initial conditions $x_1^0=6$, $x_2^0=5$, $x_3^0=4$, $x_4^0=3$, $x_j^0=2$ $(j\ge 5)$ and $y_1^t=2$ $(t\ge 0)$.

This example suggests that the equation \eqref{dKdV2} should be understood as evolving over the f\/ield~$\F_q(\delta)$, the rational function f\/ield with indeterminate $\delta$ over $\F_q$.
To obtain the time evolution pattern over $\P\F_q$, we have to substitute $\delta$ with a suitable value $\delta_0 \in\F_q$ ($\delta_0=1$ in the example above).
This substitution can be expressed as the following reduction map:
\[
\F_q(\delta)^{\times}\rightarrow \P\F_q:\ (\delta-\delta_0)^s\frac{g(\delta-\delta_0)}{f(\delta-\delta_0)}\mapsto
\begin{cases}
0,  & s>0, \\
\infty, & s<0, \\
\dfrac{g(0)}{f(0)}, & s=0,
\end{cases}
\]
where $s\in \Z$, $f(h),g(h)\in\F_q[h]$ are co-prime polynomials and $f(0)\neq 0$, $g(0)\neq 0$.
With this prescription, we know that $0/0$ does not appear and we can uniquely determine the time evolution for generic initial conditions def\/ined over $\F_q$.

\subsection{Generalized discrete KdV equation}

In this subsection we explain how to apply our method to a dynamical system with more than one parameters by taking a generalized form of the discrete KdV equation as an example.
In this case, we have to be careful in substituting the values to the parameters.
The generalised discrete KdV equation is the following discrete integrable system:
\begin{gather}
x_n^{t+1} =\dfrac{\left\{(1-\beta)+\beta x_n^ty_n^t\right\}y_n^t}{(1-\alpha)+\alpha x_n^ty_n^t},\qquad
y_{n+1}^{t} =\dfrac{\left\{(1-\alpha)+\alpha x_n^ty_n^t\right\}x_n^t}{(1-\beta)+\beta x_n^ty_n^t},
\label{YBdKdV}
\end{gather}
with arbitrary parameters $\alpha$ and $\beta$.
To avoid indeterminacy, we regard \eqref{YBdKdV} as a dynamical system over~$\F_q(\alpha, \beta)$.
Then, as in the case of~\eqref{dKdV2}, its time evolution is uniquely determined for generic initial and boundary conditions.
Note that when we substitute values in~$\F_q$ for the parameters, the result can be indeterminate, i.e., $\frac{0}{0}$, or it can depend on the order of the substitutions.
These problems are typical of a f\/ield of rational functions with two or more parameters. Even if the numerator and the denominator of a rational function are both irreducible polynomials without common factors, there will be points of indeterminacy, since the numerator and the denominator intersect in co-dimension two or more in the space of parameters.
For example, let $q=5$ and suppose that $x_n^t=y_n^t=2 \in \F_5$ then
\[
x_n^{t+1}=\frac{2+\beta}{1+3\alpha}\in \F_5(\alpha,\beta). 
\]
If we put $\alpha=\beta=3 \in \F_5$, then, we f\/ind that
\[
x_n^{t+1}=\frac{0}{0}, 
\]
or, if we f\/irst substitute $\beta$, then
\[
x_n^{t+1}=\frac{0}{1+3\alpha}\equiv 0,
\]
which is unrelated to subsequent substitutions of $\alpha$.

\begin{figure}[t]
\centering
\includegraphics[width=10cm]{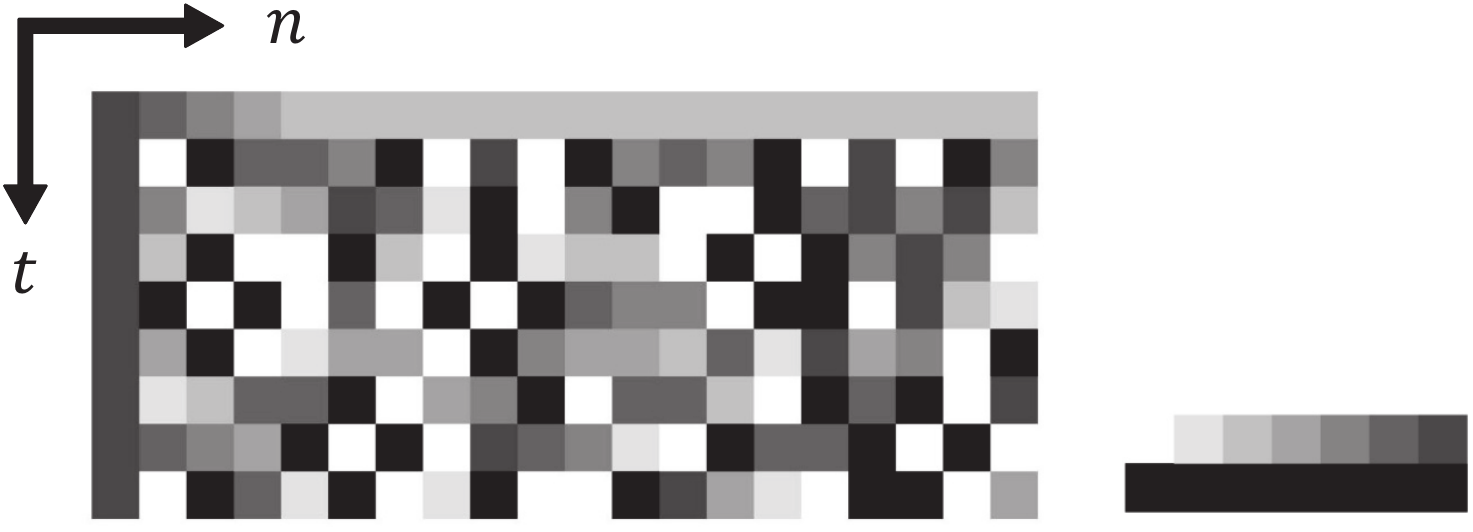}
\caption{The time evolution pattern of~$x_n^t$ of the generalised discrete KdV equation \eqref{YBdKdV} over $\P\F_7$ where $\alpha=2,\ \beta=3$. Elements of $\P\F_7$ are represented on the following grayscale: from $0$ (white) to $6$ (gray) and $\infty$ (black).}
\label{figure3}
\end{figure}

One remedy is to regard these parameters themselves as depending on a common parameter.
For example, if we put $\alpha=3+\epsilon$ and $\beta=3+\epsilon$,
\[
x_n^{t+1}=\frac{2+(3+\epsilon)}{1+3(3+\epsilon)}\equiv \frac{\epsilon}{3\epsilon}= \frac{1}{3}\equiv 2,
\]
and the value is uniquely determined in $\P\F_5$.
We show an example of a time evolution pattern of \eqref{YBdKdV} thus determined in Fig.~\ref{figure3}.
The preceding arguments suggest a general trick to construct an equation, or a time evolution rule, over a f\/inite f\/ield
from a given discrete equation.
\begin{enumerate}\itemsep=0pt
\item Introduce one parameter, say $\epsilon$, in the equation (or the initial condition), and obtain a~solution over~$\F_q(\epsilon)$.
\item Substitute a value in $\F_q$ for the parameter in the solution, and obtain a pattern over~$\P\F_q$.
\end{enumerate}
This construction can be applied to both ordinary and partial dif\/ference equations regardless of their integrability.
If we have explicit form of a solution with this parameter, we immediately obtain a pattern over a f\/inite f\/ield by replacing the parameter with a value in the f\/ield.
In the next section, we show some example of soliton solutions of~\eqref{dKdV2} and~\eqref{YBdKdV} over f\/inite f\/ields.

\section{Soliton solutions over f\/inite f\/ields}\label{section3}

\begin{figure}[t]
\centering
\includegraphics[width=10cm]{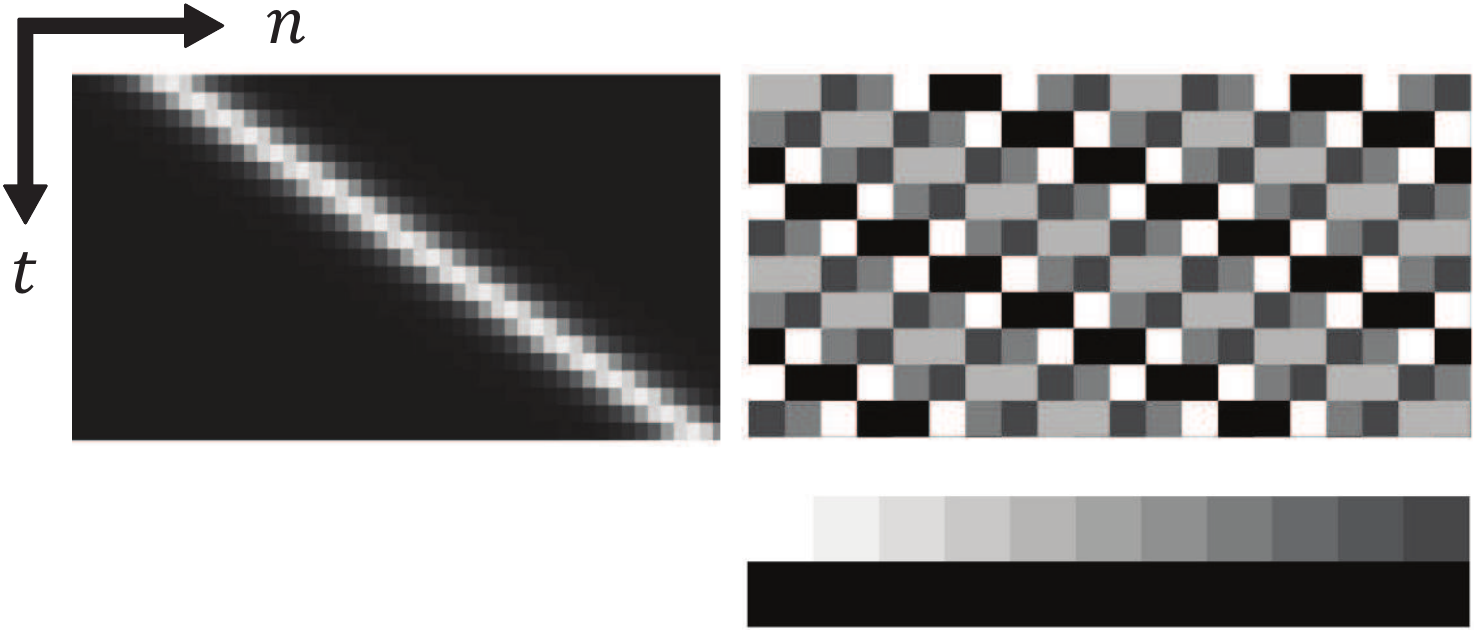}
\caption{The one-soliton solution of the discrete KdV equation \eqref{dKdV1} over $\R$ (left) and $\P\F_{11}$ (right) where $\delta=7$, $\gamma_1=2$,  $l_1=9$. Elements of~$\P\F_{11}$ are represented on the following grayscale: from $0$ (white) to $10$ (gray) and $\infty$ (black).}
\label{figure4}
\end{figure}

\begin{figure}[t!]
\centering
\includegraphics[width=10cm]{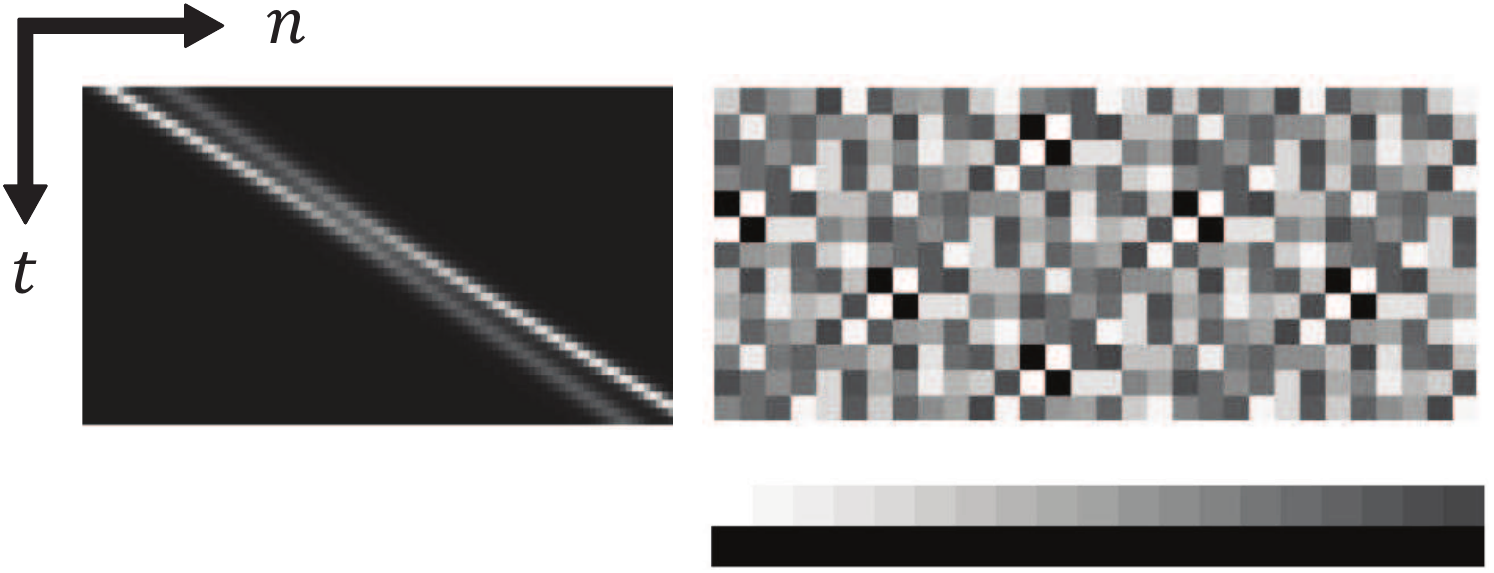}
\caption{The two-soliton solution of the discrete KdV equation \eqref{dKdV1} over $\R$ (left) and $\P\F_{19}$ (right) where $\delta=8$, $\gamma_1=15$, $l_1=2$, $\gamma_2=9$, $l_2=4$. Elements of~$\P\F_{19}$ are represented on the following grayscale: from~$0$ (white) to~$18$ (gray) and~$\infty$ (black). It is dif\/f\/icult to see the interaction of solitons over~$\P\F_{19}$.}
\label{figure5}
\end{figure}

First we consider the $N$-soliton solutions to \eqref{dKdV1} over $\F_q$.
As mentioned in the introduction, the $N$-soliton solution is given as
\begin{gather*}
x_n^t =\frac{\sigma_n^t\sigma_{n+1}^{t-1}}{\sigma_{n+1}^t\sigma_n^{t-1}},\qquad
\sigma_n^t :=\det_{1\le i,j\le N}\left( \delta_{ij}+\frac{\gamma_i}{l_i+l_j-1}\left(\frac{1-l_i}{l_i}\right)^t
\left(\frac{l_i+\delta}{1+\delta-l_i}\right)^n\right),
\end{gather*}
where $\gamma_i$, $l_i$ $(i=1,2,\dots ,N)$ are arbitrary parameters but $l_i \ne l_j$ for $i \ne j$.
When $l_i$, $\gamma_i$ are chosen in $\F_q$, $x_n^t$ becomes a rational function in~$\F_q(\delta)$.
Hence we obtain soliton solutions over~$\P\F_q$ by substituting $\delta$ with a value in~$\F_q$.
Figs.~\ref{figure4} and~\ref{figure5} show one and two soliton solutions for the discrete KdV equation \eqref{dKdV1} over the f\/inite f\/ields $\P\F_{11}$ and $\P\F_{19}$. The corresponding time evolutionary patterns on the f\/ield $\R$ are also presented for comparison.

Next we consider soliton solutions to the generalized discrete KdV equation~\eqref{YBdKdV}.
Note that
by putting $u_n^t:=\alpha x_n^t$, $v_n^t:=\beta y_n^t$, we obtain
\[
u_n^{t+1} =\dfrac{(\alpha(1-\beta)+u_n^tv_n^t)v_n^t}{\beta(1-\alpha)+u_n^t v_n^t},\qquad
v_{n+1}^{t} =\dfrac{(\beta(1-\alpha)+u_n^tv_n^t)u_n^t}{\alpha(1-\beta)+ u_n^t v_n^t}.
\]
Hence \eqref{YBdKdV} is
essentially equivalent to the `consistency of the discrete potential KdV equation around a $3$-cube' \cite{Tongasetal}: $(u,v) \to (u',v')$, as
\[
u'=vP,\qquad v'=uP^{-1},\qquad P=\dfrac{a+uv}{b+uv}.
\]
The map is also obtained from discrete BKP equation \cite{KakeiNimmoWillox}.
We will obtain $N$-soliton solutions to~\eqref{YBdKdV} from the $N$-soliton solutions to the discrete KP equation
by a reduction similar to the one adopted in~\cite{KakeiNimmoWillox}.

Let us consider the four-component discrete KP equation:
\begin{gather}
 (a_1-b)\tau_{l_1t}\tau_n+(b-c)\tau_{l_1}\tau_{tn}+(c-a_1)\tau_{l_1n}\tau_t=0,
\label{eq1}\\
 (a_2-b)\tau_{l_2t}\tau_n+(b-c)\tau_{l_2}\tau_{tn}+(c-a_2)\tau_{l_2n}\tau_t=0.
\label{eq2}
\end{gather}
Here $\tau=\tau(l_1,l_2,t,n)$ $((l_1,l_2,t,n) \in \Z^4)$ is the $\tau$-function, and $a_1$, $a_2$, $b$, $c$ are arbitrary parameters
 and we use the abbreviated form,
$\tau \equiv \tau(l_1,l_2,t,n)$, $\tau_{l_1} \equiv \tau(l_1+1,l_2,t,n)$, $\tau_{l_1t} \equiv \tau(l_1+1,l_2,t+1,n)$
and so on.
If we shift $l_1 \to l_1+1$ in \eqref{eq2}, we have
\begin{gather}
(a_2-b)\tau_{l_1l_2t}\tau_{l_1n}+(b-c)\tau_{l_1l_2}\tau_{l_1tn}+(c-a_2)\tau_{l_1l_2n}\tau_{l_1t}=0.\label{kptaushift}
\end{gather}
Then, by imposing the reduction condition:
\begin{gather}
\tau_{l_1l_2}=\tau,
\label{reduction}
\end{gather}
the equation \eqref{kptaushift} turns to
\[
(a_2-b)\tau_{t}\tau_{l_1n}+(b-c)\tau\tau_{l_1tn}+(c-a_2)\tau_{n}\tau_{l_1t}=0.
\]
Hence, putting $f:=\tau$, $g:=\tau_{l_1}$, we obtain
\begin{gather*}
\begin{split}
 (a_1-b)g_{t}f_n+(b-c)gf_{tn}+(c-a_1)g_{n}f_t=0,
\\
 (a_2-b)f_{t}g_{n}+(b-c)fg_{tn}+(c-a_2)f_{n}g_{t}=0,
 \end{split}
\end{gather*}
and
\begin{gather*}
\frac{fg_{tn}}{g f_{tn}} =\frac{(a_2-b)f_{t}g_{n}+(c-a_2) f_{n}g_{t}}{(a_1-b)g_{t}f_n+(c-a_1) g_{n}f_t}
 =\frac{(c-a_2)+(a_2-b)\frac{f_{t}g_{n}}{f_{n}g_{t}}}{(a_1-b)+(c-a_1) \frac{g_{n}f_t}{g_{t}f_n}}.
\end{gather*}
Now we denote
\begin{gather}
x_n^t:=\frac{f g_n}{g f_n},\qquad y_n^t:=\frac{g f_t}{f g_t}.
\label{xy}
\end{gather}
From the equality
\[
x_n^{t+1}y_{n+1}^t=x_n^ty_n^t=\frac{f_tg_n}{f_ng_t}, \qquad
\frac{x_n^{t+1}}{y_n^t}=\frac{f g_{tn}}{g f_{tn}},
\]
if we def\/ine  $\alpha:=\frac{c-a_1}{c-b}$, $\beta:=\frac{a_2-b}{c-b}$,
we f\/ind that $x_n^t$, $y_n^t$ def\/ined in \eqref{xy} satisfy the equation~\eqref{YBdKdV}.

The $N$-soliton solution to \eqref{eq1} and \eqref{eq2} is known as
\begin{gather*}
\tau=\det_{1\le i,j \le N}\left[ \delta_{ij}+\frac{\gamma_i}{p_i-q_j}\left(\frac{q_i-a_1}{p_i-a_1 }\right)^{l_1}
\left(\frac{q_i-a_2}{p_i-a_2 }\right)^{l_2}\left(\frac{q_i-b}{p_i-b }\right)^{t}\left(\frac{q_i-c}{p_i-c }\right)^{n} \right],
\end{gather*}
where $\{p_i,q_i\}_{i=1}^N$ are distinct parameters from each other and $\{ \gamma_i \}_{i=1}^N$ are
arbitrary parame\-ters~\cite{DJKM}.
The reduction condition \eqref{reduction} gives the constraint,
\[
\left(\frac{a_1-p_i}{a_1-q_i}\right)\left(\frac{a_2-p_i}{a_2-q_i}\right)=1,
\]
to the parameters $\{p_i,  q_i\}$.
Since $p_i \ne q_i$, the constraint becomes $p_i+q_i=a_1+a_2$.
By putting $\frac{p_i-a_1}{c-b}  \rightarrow  p_i$,
$\frac{\gamma_i}{c-b} \rightarrow  \gamma_i  $, $\Delta := \frac{a_1-a_2}{c-b}$ and $l_1=l_2$ we have
\begin{gather}
f =\det_{1\le i,j \le N}\left[ \delta_{ij}+\frac{\gamma_i}{p_i+p_j+\Delta }\left(\frac{-p_i+\beta}{p_i+1-\alpha }\right)^{t}
\left(\frac{p_i+1-\beta}{-p_i+\alpha}\right)^{n}
\right], \label{fform}\\
g =\det_{1\le i,j \le N}\left[ \delta_{ij}+\frac{\gamma_i}{p_i+p_j+\Delta }\frac{-\Delta -p_i}{p_i}\left(\frac{-p_i+\beta}{p_i+1-\alpha }\right)^{t}
\left(\frac{p_i+1-\beta}{-p_i+\alpha}\right)^{n}
\right]. \label{gform}
\end{gather}
Thus we obtain the $N$-soliton solution of \eqref{YBdKdV} by \eqref{xy}, \eqref{fform} and \eqref{gform} in the f\/ield $\F_q(\epsilon)$.

\begin{figure}[t]
\centering
\includegraphics[width=10cm]{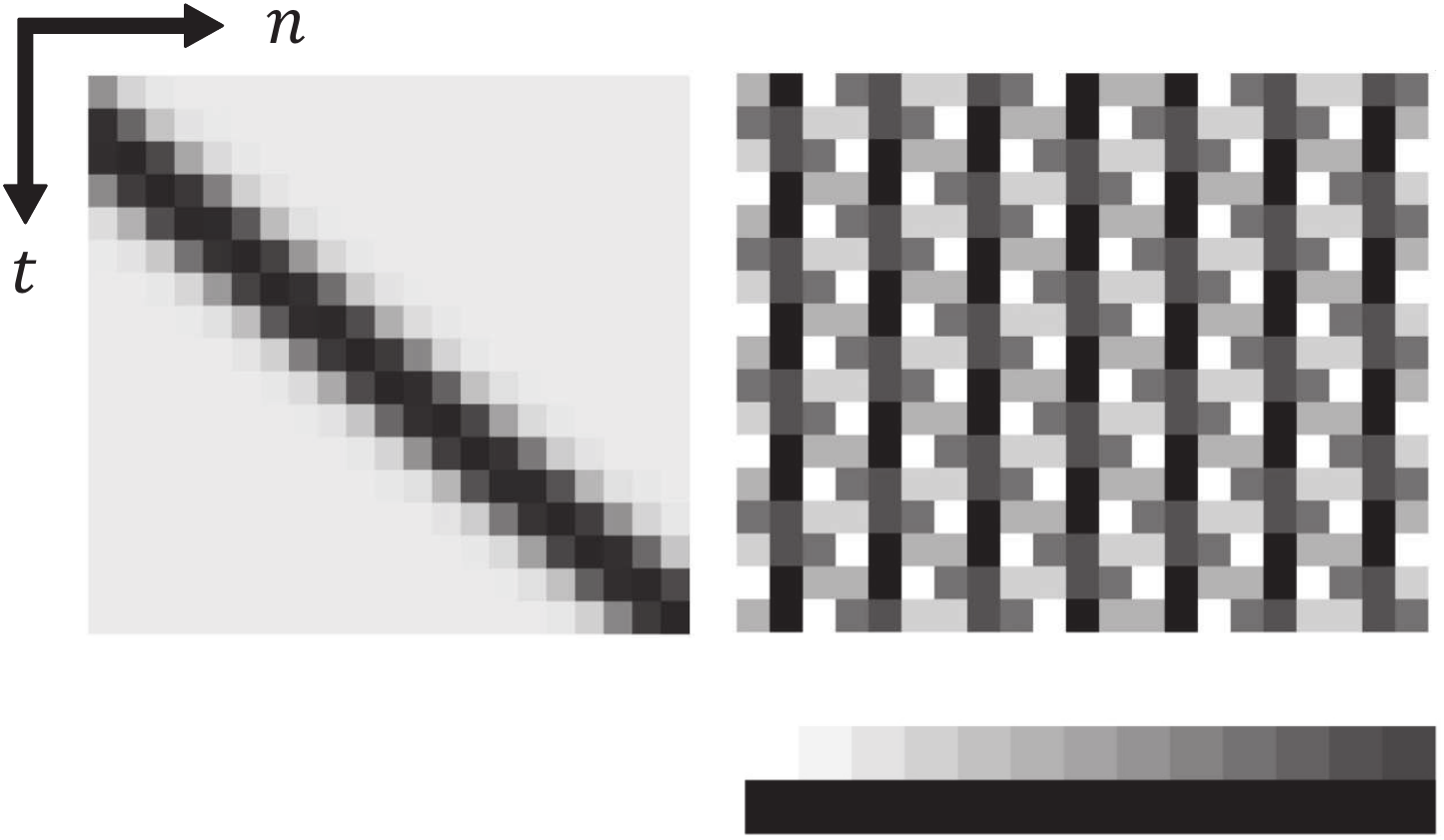}
\caption{The one-soliton solution of the generalized discrete KdV equation \eqref{YBdKdV} over $\R$ (left) and $\P\F_{13}$ (right) where $\alpha=\frac{14}{15}$, $\beta=\frac{5}{6}$, $r_1=-\frac{1}{15}$, $l_1=\frac{1}{30}$. Elements of~$\P\F_{13}$ are represented on the following grayscale: from~$0$ (white) to~$12$ (gray) and $\infty$ (black).}
\label{figure6}
\end{figure}

\begin{figure}[t!]
\centering
\includegraphics[width=10cm]{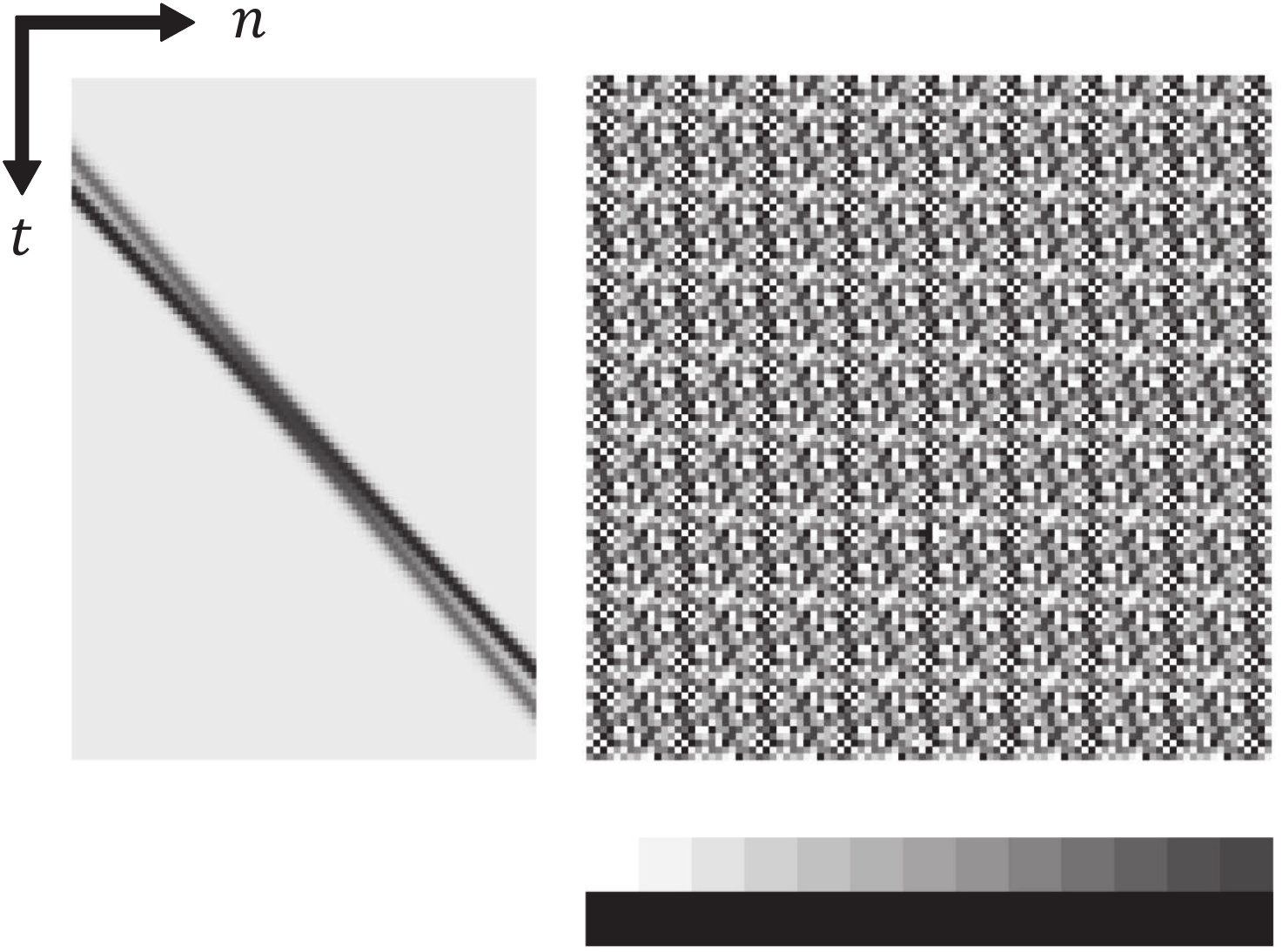}
\caption{The two-soliton solution of the generalized discrete KdV equation \eqref{YBdKdV} over $\R$ (left) and $\P\F_{13}$ (right) where $\alpha=\frac{14}{15}$, $\beta=\frac{5}{6}$, $r_1=-\frac{1}{6}$, $l_1=\frac{2}{15}$, $r_2=-\frac{1}{30}$, $l_2=\frac{1}{30}$. Elements of $\P\F_{13}$ are represented on the following grayscale: from~$0$ (white) to~$12$ (gray) and $\infty$ (black).}
\label{figure7}
\end{figure}

\begin{figure}[t]
\centering
\includegraphics[width=10cm]{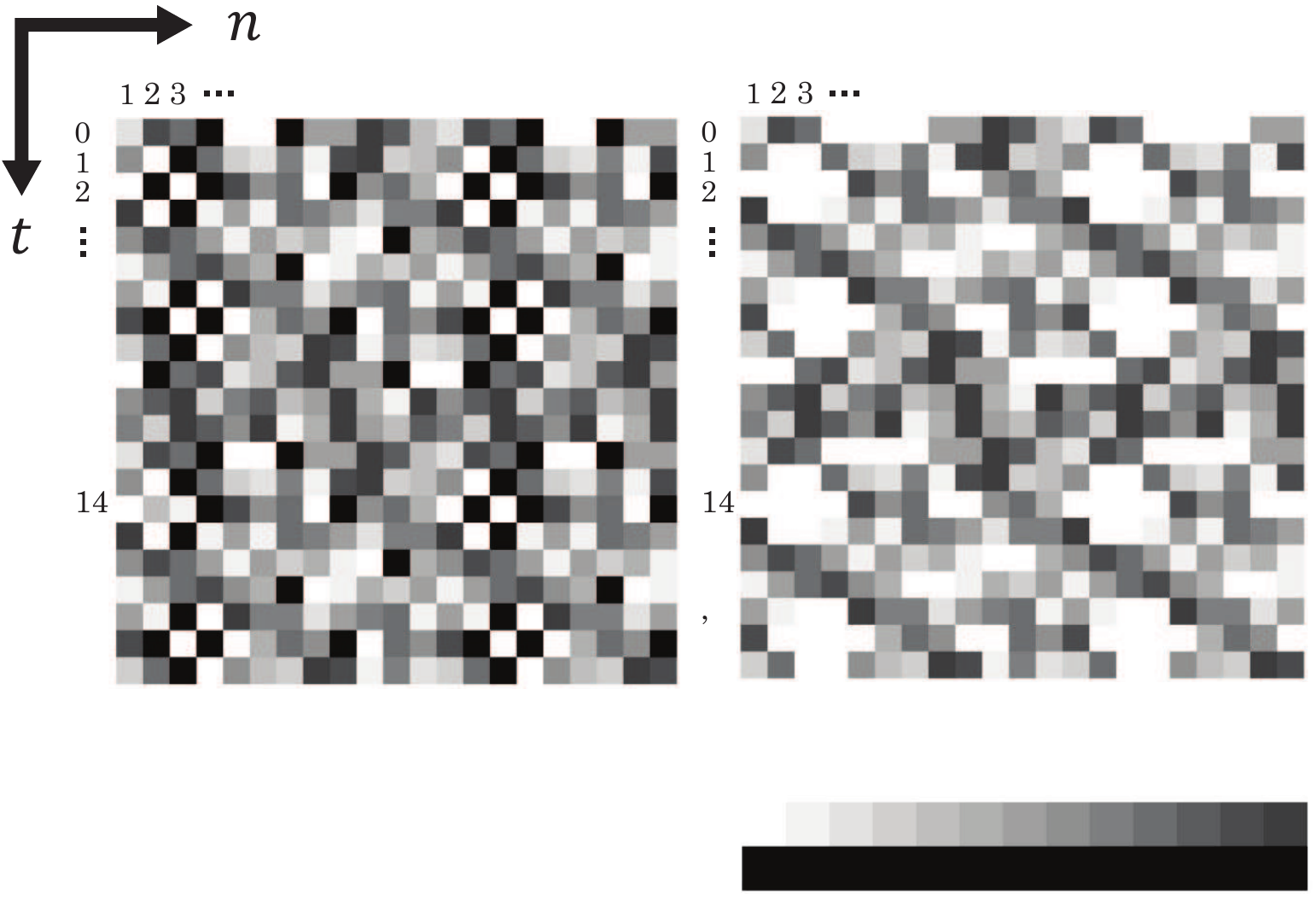}
\caption{The two-soliton solution of the generalized discrete KdV equation \eqref{YBdKdV} over~$\P\F_{13}$ calculated in two dif\/ferent ways. Elements of~$\P\F_{13}$ are represented on the following grayscale: from $0$ (white) to $12$ (gray) and $\infty$ (black).}
\label{figure8}
\end{figure}

We now return to the method of previous section.
By substituting $\alpha=n_a+\epsilon,\ \beta=n_b+\epsilon$ $(n_a, n_b \in \F_q)$, we can construct soliton solutions in $\F_q(\epsilon)$ for suitable values of
$\{p_i,  \gamma_i\}$ and $\Delta$. We denote the solutions def\/ined in $\P\F_q$ when we put $\epsilon=0$ as $\tilde{f}$, $\tilde{g}$, $\tilde{x}_n^t$ and~$\tilde{y}_n^t$.
Figs.~\ref{figure6} and~\ref{figure7} show~$\tilde{x}_n^t$ for one and two soliton solutions for the generalized discrete KdV equation \eqref{YBdKdV}.
Lastly, we discuss the periodicity of the soliton solutions over $\P\F_q$.
We have
\begin{gather*}
\tilde{f}(n+q-1, t) = \tilde{f}(n, t+q-1)=\tilde{f}(n, t),\qquad
\tilde{g}(n+q-1, t) = \tilde{g}(n, t+q-1)=\tilde{g}(n, t),
\end{gather*}
for all $t,n\in\Z$ since we have $a^{q-1}\equiv 1$ for all $a \in \F^{\times}_q$.
Thus the functions $\tilde{f}$ and $\tilde{g}$ have periods $q-1$ over $\F_q$.
However we cannot conclude that $\tilde{x}_n^t$ and $\tilde{y}_n^t$ are also periodic with periods $q-1$.
The values of $\tilde{x}_n^t$ may not be periodic when $\tilde{f}(n,t)\tilde{g}(n+1,t)=0$ and $\tilde{g}(n,t)\tilde{f}(n+1,t)=0$ (see~\eqref{xy}).
First we write $f(n,t)g(n+1,t)$ and $g(n,t)f(n+1,t)$ as follows:
\begin{gather*}
f(n,t)g(n+1,t) = \epsilon^l k(\epsilon),\qquad
g(n,t)f(n+1,t) = \epsilon^m h(\epsilon),
\end{gather*}
where $l, m\in\Z$, $h(0)\neq 0$, $k(0)\neq 0$ and $k(\epsilon),  h(\epsilon)\in \F_q[\epsilon]$.
We also write $f(n+q-1,t)g(n+q,t)=\epsilon^{l'} k'(\epsilon)$, $g(n+q-1,t)f(n+q,t)=\epsilon^{m'} h'(\epsilon)$ in the same manner.
Let us write down the reduction map again:
\[
\tilde{x}_n^t=
\begin{cases}
\dfrac{k(0)}{h(0)},  & l=m, \\
0, & l>m,\\
\infty, & l<m.
\end{cases}
\]
In the case when $\tilde{f}(n,t)\tilde{g}(n+1,t)=0$ and $\tilde{g}(n,t)\tilde{f}(n+1,t)=0$, $x_n^t=\frac{f(n,t)g(n+1,t)}{g(n,t)f(n+1,t)}\in\F_q(\epsilon)$ and $x_n^{t+q-1}=\frac{f(n+q-1,t)g(n+q,t)}{g(n+q-1,t)f(n+q,t)}\in\F_q(\epsilon)$ can have dif\/ferent reductions with respect to $\epsilon$, since $l'$ is not necessarily equal to $l$, and neither $m'$ is equal to $m$.
The left part of Fig.~\ref{figure8} shows a magnif\/ied plot of the same two-soliton solutions as in Fig.~\ref{figure7}. In some points $\tilde
{x}_n^t$ does not have~period 12 (for example~$x_2^2\neq x_2^{14}$) while almost all other points do have this periodicity.

If we want to recover full periodicity, there is another reduction to obtain $\tilde{x}_n^t$ and $\tilde{y}_n^t$ from $x_n^t$ and $y_n^t$.
This time, we def\/ine $\tilde{x}_n^t$ as
\[
\tilde{x}_n^t=
\begin{cases}
\dfrac{k(0)}{h(0)},  & l=0,\ m=0, \\
0, & \mbox{otherwise}.
\end{cases}
\]
The right part of Fig.~\ref{figure8} shows the same two-soliton solution as in the left part but calculated with this new method. We see that all points have periods~12.
It is important to determine how to reduce values in~$\P\F_q(\epsilon)$ to values in~$\P\F_q$, depending on the properties one wishes the soliton solutions to possess.

\section{Discussion and concluding remarks}\label{section4}

We presented a prescription to obtain dynamical equations over f\/inite f\/ields from discrete equations def\/ined over a f\/ield with characteristic~$0$.
The essential trick is to introduce an indeterminate (a parameter~$\epsilon$) and regard the equations def\/ined over $\F_q(\epsilon)$.
By substituting $\epsilon$ with a value in~$\F_q$, we can uniquely determine the values of the dependent variables in~$\P\F_q$.
Furthermore if we have a one-parameter family of solutions to the original equation, we obtain the solution over a f\/inite f\/ield
straight away by substituting a suitable value for the parameter.
The $N$-soliton solutions to the discrete KdV equation and the generalized discrete KdV equations over f\/inite f\/ields are thus obtained.

Our approach is also applied not only to the other discrete soliton equations but to ordinary nonlinear dif\/ference equations such as discrete Painlev\'{e} equations.
Let us consider dP2 equa\-tion~\cite{NP,RGH}:
\begin{gather}
x_{n+1}+x_{n-1}=\frac{z_nx_n+a}{x_n^2-1}, \qquad n \in \Z,
\label{dP2}
\end{gather}
where $z_n:=n \delta$ and $a$, $\delta$ are constants.
If we examine~\eqref{dP2} over~$\F_q$, we cannot def\/ine its time evolution after the dependent variable $x_n$ takes $\pm 1$.
However, by regarding $a$ as an indeterminate, we can def\/ine the time evolution over~$\F_q(a)$ and obtain a value in $\P\F_q$ by substituting a value in $\F_q$ for~$a$.
The ef\/fectiveness of this approach is conf\/irmed for wide range of ordinary discrete equations such as the other discrete or $q$-discrete Painlev\'{e} equations and the QRT map\-pings~\cite{Tamizhmani}.

There is, however, another choice of indeterminate for the discrete ordinary dif\/ference equations.
Let $(x_0,x_1)=(y, x)$ with $y \in \F_q$ in \eqref{dP2}.
Then $x_k$ $(k=2,3,\dots )$ can be regarded as a~function of $x$, i.e.~$x_k \in \F_q(x)$.
Since we have
\begin{gather*}
x_2=\frac{z_1+a}{2(x-1)}+\frac{(z_1-a)}{4}+O(x-1), \\
x_3=-1+O(x-1), \\
x_4=\frac{y(z_1+a)+2a+\delta z_2}{a-z_3}+O(x-1),\\
\cdots\cdots\cdots\cdots,
\end{gather*}
we can determine $x_3,x_4,\dots $ by putting $ x \to 1$ despite the fact that they are not well def\/ined if we take $x_1=1$ in advance.
The time evolution pattern thus obtained coincides with that of our approach with an indeterminate~$a$.

The above procedure reminds us the singularity conf\/inement method which is an ef\/fective test to judge the integrability of the given equations~\cite{Grammaticosetal}.
In fact, if we consider \eqref{dP2} over $\C$ and take $\epsilon:=x-1$ as an inf\/initesimal parameter,
the time evolution pattern is exactly the one which passes the singularity conf\/inement test.
The grounds of this similarity become clear when one thinks of the theory of the space of initial conditions of the Painlev\'{e} equations~\cite{Okamoto1,Okamoto2,Okamoto3,Okamoto4,Sakai}.
As observed by Sakai~\cite{Sakai}, passing the singularity conf\/inement test is essentially equivalent to the fact that the equation is lifted to an automorphism of the rational surface (the space of initial conditions) obtained by compactif\/ication and blowing-up from the original space of initial values $\C \times \C$.
By introducing inf\/initesimal parameter in the case of singularity conf\/inement test or an indeterminate in our approach, we avoid passing through a point on an exceptional curve generated by blowing-up and approximate the automorphism in an ef\/fective way.
Hence these three types of approaches, that is, construction of space of initial conditions, application of singularity conf\/inement and the method shown in the present paper,
give the same time evolution rule for the discrete Painelev\'{e} equations over f\/inite f\/ields.

For soliton equations, however, we have dif\/f\/iculty in def\/ining time evolution by constructing the space of initial conditions.
Let us return to the equation~\eqref{YBdKdV}.
The mapping, $(x_n^t,y_n^t) \mapsto (x_n^{t+1}, y_{n+1}^t)$, is lifted to an automorphism of the surface~$\tilde{X}$,
where~$\tilde{X}$ is obtained from $\P^1 \times \P^1$ by blowing up twice at $(0,\infty)$ and $(\infty, 0)$ respectively:
\begin{gather*}
\tilde{X} =\A_1 \cup \A_2,\\
\A_1 :=\left\{ \left(\left(x, \frac{1}{y}\right), [\xi:\eta], [u:v]\right)  \Big| \ x \eta-\frac{1}{y} \xi=0,\right. \\
 \hphantom{\A_1 :=}{} \qquad  \quad x\left((1-\alpha)\eta+\alpha \xi\right)v-\left((1-\beta)\eta+\beta \xi\right)u=0 \biggr\}\subset \A^2 \times \P^1\times\P^1, \\
\A_2 :=\left\{ \left(\left(\frac{1}{x}, y\right), [\xi:\eta], [w:z]\right)  \Big|\ \frac{1}{x} \xi-y \eta=0, \right. \\
\hphantom{\A_2 :=}{}
\qquad  \quad y\left((1-\beta)\eta+\beta \xi\right)w-\left((1-\alpha)\eta+\alpha \xi\right)z=0 \biggr\}\subset \A^2 \times \P^1\times\P^1,
\end{gather*}
where $[a:b]$ denotes a set of homogeneous coordinates for $\P^1$.
But, to def\/ine the time evolution of the system with $N$ lattice points  from \eqref{YBdKdV}, we have to consider the mapping
\[
(y_1^t;x_1^t,x_2^t,\dots ,x_N^t) \longmapsto  \big(x_1^{t+1},x_2^{t+1},\dots ,x_N^{t+1};y_{N+1}^t\big).
\]
Since there seems no reasonable decomposition of $\tilde{X}$ into a direct product of two independent spaces, successive use of~\eqref{YBdKdV} becomes impossible.
Note that if we blow down $\tilde{X}$ to $\P^1 \times \P^1$, the information of the initial values is lost in general.
If we intend to construct an automorphism of a space of initial conditions, it will be inevitable to start from~$\P^{N+1}$ and blow-up to some huge manifold, which is beyond the scope of the present paper.
This dif\/f\/iculty seems to be one of the reasons why the singularity conf\/inement method has not been used for construction of integrable partial dif\/ference equations or judgement for their integrability, though some attempts have been proposed in the bilinear form~\cite{RGS}.
There should be so many exceptional hyperplanes in the space of initial conditions if it does exist, and it is practically impossible to check all the ``singular'' patterns in the na\"{i}ve extension of the singularity conf\/inement test.
On the other hand, when we f\/ix the initial condition for a partial dif\/ference equation, the number of singular patterns is restricted in general and we have only to enlarge the domain so that the mapping becomes well def\/ined.
This is the strategy that we adopted in this article.

As shown in the above discussion, the discrete Painlev\'{e} equations over f\/inite f\/ields can be treated by several methods.
The comparison of the mathematical structure in~$\C$ with that in~$\F_q$ is one of the future problems.
Clarifying the geometric and/or algebraic meaning of our approach to soliton equations and applications of our approach to  the initial value problems related to curves over f\/inite f\/ields are also the problems we shall address in the future.

\subsection*{Acknowledgement}
The authors wish to thank Professors K.M.~Tamizhmani, R.~Willox and Dr. S.~Iwao for useful comments.  This work is partially supported by Grant-in-Aid for JSPS Fellows (24-1379).

\pdfbookmark[1]{References}{ref}
\LastPageEnding


\begin{thebibliography}{99}
\footnotesize\itemsep=0pt

\bibitem{BD}
Bia{\l}ecki M., Doliwa A., Discrete {K}adomtsev--{P}etviashvili and
  {K}orteweg--de {V}ries equations over f\/inite f\/ields, \href{http://dx.doi.org/10.1023/A:1026000605865}{\textit{Theoret. and
  Math. Phys.}} \textbf{137} (2003), 1412--1418, \href{http://arxiv.org/abs/nlin.SI/0302064}{nlin.SI/0302064}.

\bibitem{BN}
Bia{\l}ecki M., Nimmo J.J.C., On pattern structures of the {$N$}-soliton
  solution of the discrete {KP} equation over a f\/inite f\/ield,
  \href{http://dx.doi.org/10.1088/1751-8113/40/5/006}{\textit{J.~Phys.~A: Math. Theor.}} \textbf{40} (2007), 949--959,
  \href{http://arxiv.org/abs/nlin.SI/0608041}{nlin.SI/0608041}.

\bibitem{DJKM}
Date E., Jinbo M., Miwa T., Method for generating discrete soliton
  equation.~II, \href{http://dx.doi.org/10.1143/JPSJ.51.4125}{\textit{J.~Phys. Soc. Japan}} \textbf{51} (1982), 4125--4131.

\bibitem{DBK}
Doliwa A., Bia{\l}ecki M., Klimczewski P., The {H}irota equation over f\/inite
  f\/ields: algebro-geometric approach and multisoliton solutions,
  \href{http://dx.doi.org/10.1088/0305-4470/36/17/309}{\textit{J.~Phys.~A: Math. Gen.}} \textbf{36} (2003), 4827--4839,
  \href{http://arxiv.org/abs/nlin.SI/0211043}{nlin.SI/0211043}.

\bibitem{Grammaticosetal}
Grammaticos B., Ramani A., Papageorgiou V., Do integrable mappings have the
  {P}ainlev\'e property?, \href{http://dx.doi.org/10.1103/PhysRevLett.67.1825}{\textit{Phys. Rev. Lett.}} \textbf{67} (1991),
  1825--1828.

\bibitem{KakeiNimmoWillox}
Kakei S., Nimmo J.J.C., Willox R., Yang--{B}axter maps and the discrete {KP}
  hierarchy, \href{http://dx.doi.org/10.1017/S0017089508004825}{\textit{Glasg. Math.~J.}} \textbf{51} (2009), 107--119.

\bibitem{NP}
Nijhof\/f F.W., Papageorgiou V.G., Similarity reductions of integrable lattices
  and discrete analogues of the {P}ainlev\'e {${\rm II}$} equation,
  \href{http://dx.doi.org/10.1016/0375-9601(91)90955-8}{\textit{Phys. Lett.~A}} \textbf{153} (1991), 337--344.


\bibitem{Okamoto1}
Okamoto K., Studies on the {P}ainlev\'e equations. {I}.~{S}ixth {P}ainlev\'e equation {$P_{{\rm VI}}$},
\href{http://dx.doi.org/10.1007/BF01762370}{\textit{Ann. Mat. Pura Appl.~(4)}}
\textbf{146} (1987), 337--381.

\bibitem{Okamoto2}
Okamoto K., Studies on the {P}ainlev\'e equations. {II}.~{F}ifth {P}ainlev\'e equation {$P_{\rm V}$},
\textit{Japan.~J. Math.~(N.S.)} \textbf{13} (1987), 47--76.

\bibitem{Okamoto3}
Okamoto K., Studies on the {P}ainlev\'e equations. {III}.~{S}econd and fourth {P}ainlev\'e equations, {$P_{{\rm II}}$} and {$P_{{\rm IV}}$},
\href{http://dx.doi.org/10.1007/BF01458459}{\textit{Math. Ann.}} \textbf{275} (1986), 221--255.

\bibitem{Okamoto4}
Okamoto K., Studies on the {P}ainlev\'e equations. {IV}.~{T}hird {P}ainlev\'e
  equation {$P_{{\rm III}}$}, \textit{Funkcial. Ekvac.} \textbf{30} (1987),
  305--332.

\bibitem{Tongasetal}
Papageorgiou V.G., Tongas A.G., Veselov A.P., Yang--{B}axter maps and
  symmetries of integrable equations on quad-graphs, \href{http://dx.doi.org/10.1063/1.2227641}{\textit{J.~Math. Phys.}}
  \textbf{47} (2006), 083502, 16~pages, \href{http://arxiv.org/abs/math.QA/0605206}{math.QA/0605206}.

\bibitem{RGH}
Ramani A., Grammaticos B., Hietarinta J., Discrete versions of the {P}ainlev\'e
  equations, \href{http://dx.doi.org/10.1103/PhysRevLett.67.1829}{\textit{Phys. Rev. Lett.}} \textbf{67} (1991), 1829--1832.

\bibitem{RGS}
Ramani A., Grammaticos B., Satsuma J., Integrability of multidimensional
  discrete systems, \href{http://dx.doi.org/10.1016/0375-9601(92)90235-E}{\textit{Phys. Lett.~A}} \textbf{169} (1992), 323--328.

\bibitem{Sakai}
Sakai H., Rational surfaces associated with af\/f\/ine root systems and geometry of
  the {P}ainlev\'e equations, \href{http://dx.doi.org/10.1007/s002200100446}{\textit{Comm. Math. Phys.}} \textbf{220} (2001),
  165--229.

\bibitem{Tamizhmani}
Tamizhmani K.M., {P}rivate communication, 2012.

\bibitem{Wolfram}
Wolfram S., Statistical mechanics of cellular automata, \href{http://dx.doi.org/10.1103/RevModPhys.55.601}{\textit{Rev. Modern
  Phys.}} \textbf{55} (1983), 601--644.

\end{thebibliography}
\end{document}